\documentclass[12pt,epsf,epsfig,psfig]{article}
\usepackage{graphicx}
\usepackage{epsfig}
\def\J{$J/\psi$}

\def\X{$\chi_c$}

\def\P{$\psi'$}

\def\U{$\Upsilon$}

\def\C{c{\bar c}}

\def\q{q{\bar q}}
\def\Q{Q{\bar Q}}
\def\e{\epsilon}
\def\t{\tau}
\def\l{\Lambda_{\rm QCD}}

\def\PL{{ Phys.\ Lett.\ }}
\def\PR{{ Phys.\ Rev.\ }}

\def\PRL{{ Phys.\ Rev.\ Lett.\ }}

\def\ZP{{ Z.\ Phys.\ }}
\def\EPJ{{Eur.\ Phys.\ J.\ }}

\def\be{\begin{equation}}
\def\ee{\end{equation}}
\def\lsim{\raise0.3ex\hbox{$<$\kern-0.75em\raise-1.1ex\hbox{$\sim$}}}
\def\gsim{\raise0.3ex\hbox{$>$\kern-0.75em\raise-1.1ex\hbox{$\sim$}}}

\begin{document}

BI-TP 2006/06\hfill 28.\ 2.\ 06

~\vskip 1cm

\centerline{\Large \bf Charm and Beauty in a Hot Environment$^*$} 

\vskip1cm

\centerline{\large \bf Helmut Satz} 

\vskip 0.5cm

\medskip

\centerline{Universit\"at Bielefeld, Germany}


 
\centerline{and}

\centerline{Instituto Superior T\'ecnico, Lisboa, Portugal}



\vskip1cm

\centerline{\bf Abstract:}

\vskip0.5cm

We discuss the spectral analysis of quarkonium states in a hot medium of 
deconfined quarks and gluons, and we show that such an analysis provides 
a way to determine the thermal properties of the quark-gluon plasma.

\vskip1cm

\noindent{\large \bf 1.\ Introduction}

\bigskip

Since some thirty years it is known that besides the almost massless 
{\it up} and $down$ quarks of the everyday world, and the still relatively 
light {\it strange} quarks required to account for the strange mesons and 
hyperons observed in hadron collisions, there are heavy quarks at the other 
end of the scale, whose bare masses alone are larger than those of most of 
the normal hadrons. These heavy quarks first showed up in the discovery of 
the \J ~meson \cite{Ting}, of mass of 3.1 GeV; it is a bound state of a 
{\it charm} quark ($c$) and its antiquark ($\bar c$), each having a mass 
of some 1.2--1.5 GeV. On the next level there is the \U~meson \cite{Leder}, 
with a mass of about 9.5 GeV, made up of a {\it bottom} or {\it beauty} 
quark-antiquark pair ($b \bar b$), with each quark here having a mass 
around 4.5 - 4.8 GeV. Both charm and bottom quarks can of course also 
bind with normal light quarks, giving rise to open charm ($D$) and open 
beauty ($B$) mesons. The lightest of these `light-heavy' mesons have 
masses of about 1.9 GeV and 5.3 GeV, respectively. 

\medskip

The bound states of a heavy quark $Q$ and its antiquark $\bar Q$ are 
generally referred to as quarkonia. Besides the initially discovered 
vector ground states \J~and \U, both the $c \bar c$ and the $b \bar b$ 
systems give rise to a number of other {\it stable} bound states of 
different quantum numbers. They are stable in the sense that their mass 
is less than that of two light-heavy mesons, so that strong decays are 
forbidden. The measured stable charmonium spectrum contains the $1S$ 
scalar $\eta_c$ and vector \J, three $1P$ states \X~(scalar, vector 
and tensor), and the $2S$ vector state \P, whose mass is just below 
the open charm threshold. There are further charmonium states above the 
\P; these can decay into $D \bar D$ pairs, and we shall here restrict 
our considerations only to quarkonia stable under strong interactions.

\medskip

\hrule width5cm

\medskip

$*$ Dedicated to Adriano Di Giacomo on the occasion of his 70th birthday

\newpage

The study of quarkonia has played and continues to play a major role
in many aspects of QCD: in spectroscopy, in production and decay, and
last but not least as probe of hot QCD media \cite{nora}. And in the
development of our understanding of and our feeling for the charm and
the beauty of strong interactions, Adriano Di Giacomo has played and
certainly will continue to play a major role. Neither my competence nor
the space and time available here allow me to give an overview of these
contributions; so let me just cite a few recent ones which are of
particular importance to what I want to discuss here \cite{DG,A-DG}. 

\medskip

Quarkonia are rather unusual hadrons. The masses of the light hadrons,
in particular those of the non-strange mesons and baryons, arise almost
entirely from the interaction energy of their nearly massless quark 
constituents. In contrast, the quarkonium masses are largely determined
by the bare charm and bottom quark masses. These large quark masses allow 
a very straightforward calculation of many basic quarkonium properties, 
using non-relativistic potential theory. It is found that, in particular,
the ground states and the lower excitation levels of quarkonia are very
much smaller than the normal hadrons, and that they are very tightly bound.
Now deconfinement is a matter of scales: when the separation between normal 
hadrons becomes much less than the size of these hadrons, they melt to form 
the quark-gluon plasma. What happens at this point to the much smaller 
quarkonia?  When do they become dissociated? That is the main question
we want to address here.
We shall show that the disappearance of specific quarkonia signals the 
presence of a deconfined medium of a specific temperature \cite{M-S}.
Thus the study of the quarkonium spectrum in a given medium 
is akin to the spectral analysis of stellar media, where the
presence or absence of specific excitation lines allows a determination
of the temperature of the stellar interior.

\medskip

We had defined quarkonia as bound states of heavy quarks which are 
stable under strong decay, i.e., $M_{c\bar c} \leq 2 M_D$ for charmonia 
and $M_{b\bar b} \leq 2 M_B$ for bottomonia. Since the quarks are heavy,
with $m_c \simeq 1.2 - 1.5 $ GeV for the charm and $m_b \simeq 4.5 - 4.8$ 
GeV for the bottom quark, quarkonium spectroscopy can be studied quite well 
in non-relativistic potential theory. The simplest (``Cornell'') confining
potential \cite{Cornell} for a $\Q$ at separation distance $r$ has the form
\be
V(r) = \sigma ~r - {\alpha \over r}
\label{cornell}
\ee
with a string tension $\sigma \simeq 0.2$ GeV$^2$ and a Coulomb-like term
with a gauge coupling $\alpha \simeq \pi/12$. The corresponding 
Schr\"odinger equation
\be
\left\{2m_c -{1\over m_c}\nabla^2 + V(r)\right\} \Phi_i(r) = M_i \Phi_i(r)
\label{schroedinger}
\ee
then determines the bound state masses $M_i$ and the wave functions 
$\Phi_i(r)$, and with
\be
\langle r_i^2 \rangle = \int d^3r~ r^2 |\Phi_i(r)|^2 / 
\int d^3r~|\Phi_i(r)|^2 .
\label{radii}
\ee
the latter in turn provide the (squared) average bound state ``radii'',
here defined as the $\Q$ separation for the state in question.

\medskip

The solution of eq.\ (\ref{schroedinger}) gives in fact a very good 
account of the full (spin-averaged) quarkonium spectroscopy, as seen in
Table 1 \cite{D-K-S}. The line labelled $\Delta M$ shows the differences 
between the experimental and the calculated values; they are in all cases 
less than 1 \%. The input parameters for these results were $m_c=1.25$ GeV,
$m_b=4.65$ GeV, $\sqrt \sigma = 0.445$ GeV, $\alpha=\pi/12$.
We see that in particular the \J~ and the lower-lying bottomonium
states are very tightly bound ($ 2M_{D,B} - M_0 \gg \l)$ and of very
small spatial size ($r_0 \ll 2 r_h \simeq 2$ fm). What happens to them
in a hot and dense medium?

\vskip0.5cm

\hskip1.5cm
\renewcommand{\arraystretch}{1.8}
\begin{tabular}{|c|c|c|c|c|c|c|c|c|}
\hline
{\rm state}& $J/\psi$ & $\chi_c$ & $\psi'$  & $\Upsilon$
 & $\chi_b$ & 
$\Upsilon'$ & $\chi_b'$ & $\Upsilon''$ \\
\hline
{\rm mass~[GeV]}&
3.10&
3.53&
3.68&
9.46&
9.99&
10.02&
10.26&
10.36 \\
\hline
$\Delta E$ {\rm[GeV]}&0.64&0.20&0.05 &1.10&
0.67&0.54&0.31&0.20 \cr
\hline
$\Delta M$ {\rm[GeV]}&0.02&-0.03&0.03 & 0.06&
-0.06&-0.06&-0.08&-0.07 \cr
\hline
{$r_0$ \rm [fm]}&0.50&0.72&0.90&
0.28&0.44& 0.56& 0.68 &0.78 \cr
\hline
\end{tabular}

\bigskip

\centerline{Table 1:
Quarkonium Spectroscopy from Non-Relativistic Potential Theory \cite{D-K-S}}

\vskip0.8cm

\noindent{\large \bf 2.\ Interaction Range and Colour Screening}

\bigskip

\begin{figure}[htb]
\hspace*{0.1cm}
\centerline{\epsfig{file=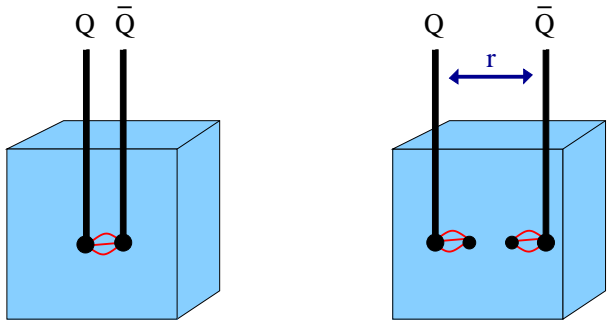,width=8cm}}
\caption{String breaking for a $\Q$ system}
\label{string-b}
\end{figure}

Consider a colour-singlet bound state of a heavy quark $Q$ and its 
antiquark $\bar Q$, put into the medium in such a way that we can measure 
the energy of the system as function of the $\Q$ separation $r$ (see Fig.\ 
\ref{string-b}). The quarks are assumed to be heavy so that they are static 
and any energy changes indicate changes in the binding energy. We consider 
first the case of vanishing baryon density; at $T=0$, the box is therefore 
empty. 

\medskip

\medskip

In vacuum, i.e., at $T=0$, the free energy of the $\Q$ pair is
assumed to have the string form \cite{Cornell} 
\be
F(r) \sim \sigma r
\label{string}
\ee
where $\sigma \simeq 0.16$ GeV$^2$ is the string tension as 
determined in the spectroscopy of heavy quark resonances (charmonium
and bottomonium states). Thus $F(r)$ increases with separation
distance; but when it reaches the value of a pair of dressed
light quarks (about the mass of a $\rho$ meson), it becomes energetically
favorable to produce a $\q$ pair from the vacuum, break the string
and form two light-heavy mesons ($Q\bar q$) and ($\bar Q q$). These
can now be separated arbitrarily far without changing the energy of the
system (Fig.\ \ref{string-b}). 

\medskip

The string breaking energy for charm quarks is found to be
\be
F_0 = 2(M_D - m_c) \simeq 1.2~{\rm GeV};
\label{charm-break}
\ee
for bottom quarks, one obtains the same value,
\be
 F_0 = 2(M_B - m_b) \simeq 1.2~{\rm GeV},
\label{bottom-break}
\ee
using in both cases the quark mass values obtained in the solution
leading to Table 3. Hence the onset of string breaking is evidently 
a property of the vacuum as a medium. It occurs when the two heavy 
quarks are separated by a distance
\be
r_0 \simeq 1.2~{\rm GeV}/\sigma \simeq 1.5~{\rm fm},
\label{stringbreak}
\ee
independent of the mass of the (heavy) quarks connected by the string.

\medskip

\begin{figure}[h]
\hspace*{0.1cm}
\centerline{\epsfig{file=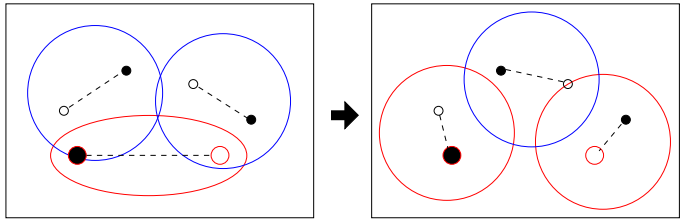,width=9cm}}
\caption{In-medium string breaking through recoupling}
\label{flip}
\end{figure}

If we heat the system to get $T>0$, the medium begins to contain light 
mesons, and the large distance $\Q$ potential $F(\infty,T)$ decreases, since 
we can use these light hadrons to achieve an earlier string breaking through 
a kind of flip-flop recoupling of quark constituents \cite{miya}, resulting 
in an 
effective screening of the interquark force (see Fig.\ \ref{flip}).
Near the deconfinement point, the hadron density increases rapidly, and
hence the recoupling dissociation becomes much more effective, causing
a considerable decrease of $F(\infty,T)$.

\medskip

A further increase of $T$ will eventually bring the medium to the 
deconfinement point $T_c$, at which chiral symmetry restoration causes 
a rather abrupt drop of the light quark dressing (equivalently, of the 
constituent quark mass), increasing strongly the density of constituents. 
As a consequence, $F(\infty,T)$ now continues to drop sharply.   
Above $T_c$, light quarks and gluons become deconfined colour charges, and 
this quark-gluon plasma leads to a colour screening, which limits the range
of the strong interaction. The colour screening radius $r_D$, which determines 
this range, is inversely proportional to the density of charges, so that
it decreases with increasing temperature. As a result, the $\Q$ interaction
becomes more and more short-ranged. 

\medskip

\begin{figure}[htb]
\hspace*{-0.3cm}
\centerline{\epsfig{file=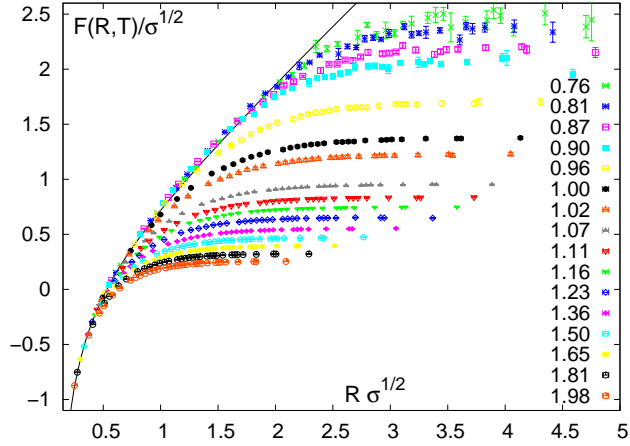,width=9cm}}
\caption{The colour singlet $\Q$ free energy $F(r,T)$ vs.\ $r$ at 
different $T$ \cite{K-Z}}
\label{OKfree}
\end{figure}

In summary, starting from $T=0$, the $\Q$ probe first tests vacuum string 
breaking, then a screening-like
dissociation through recoupling of constituent quarks, and 
finally genuine colour screening. In Fig.\ \ref{OKfree}, we show the 
behaviour obtained in full two-flavour QCD for the colour-singlet
$\Q$ free energy as a function of $r$ for different $T$ \cite{K-Z}. 

\medskip

\begin{figure}[htb]
\mbox{
\hskip1cm
\epsfig{file=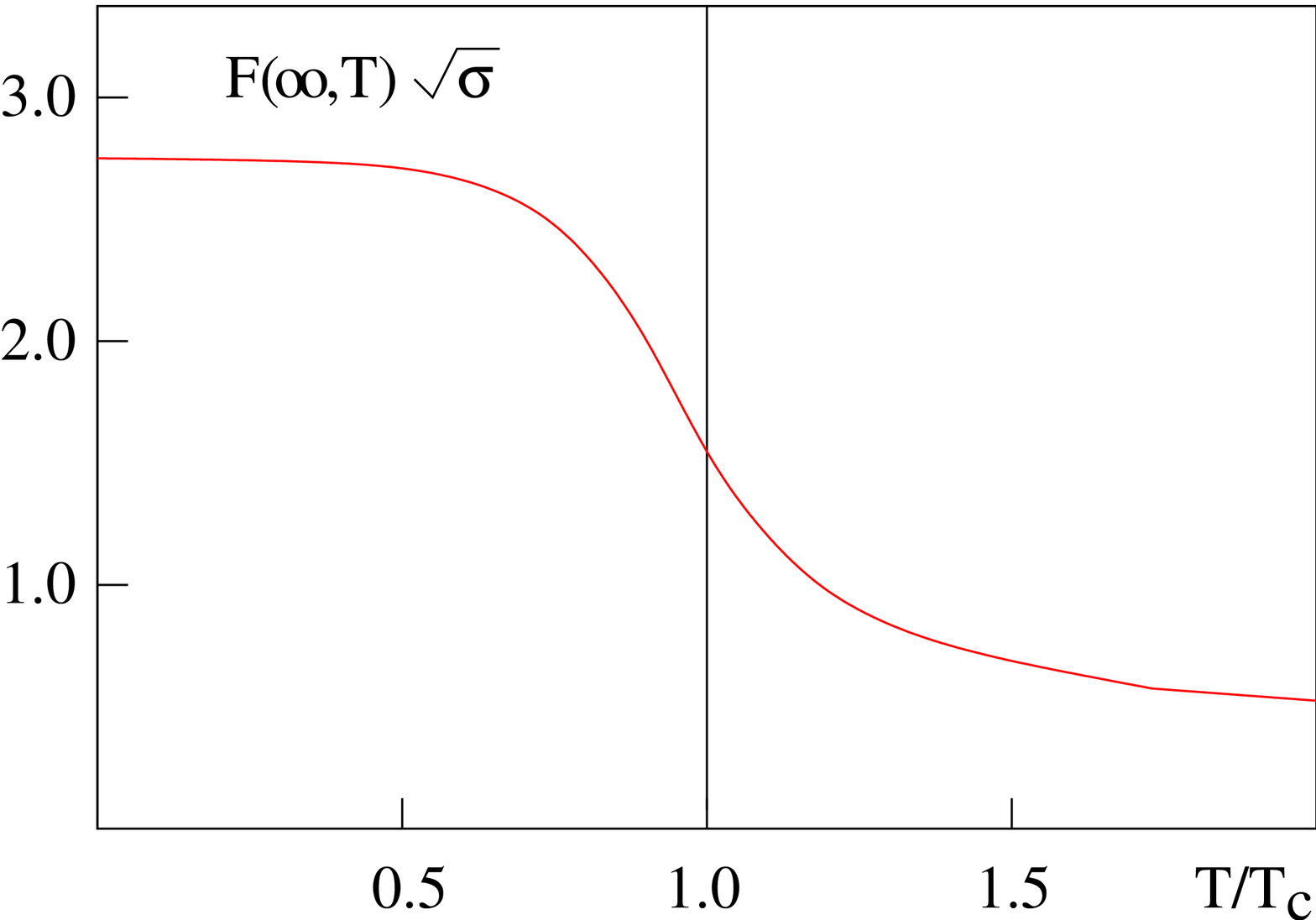,width=5cm}
\hskip3cm
\epsfig{file=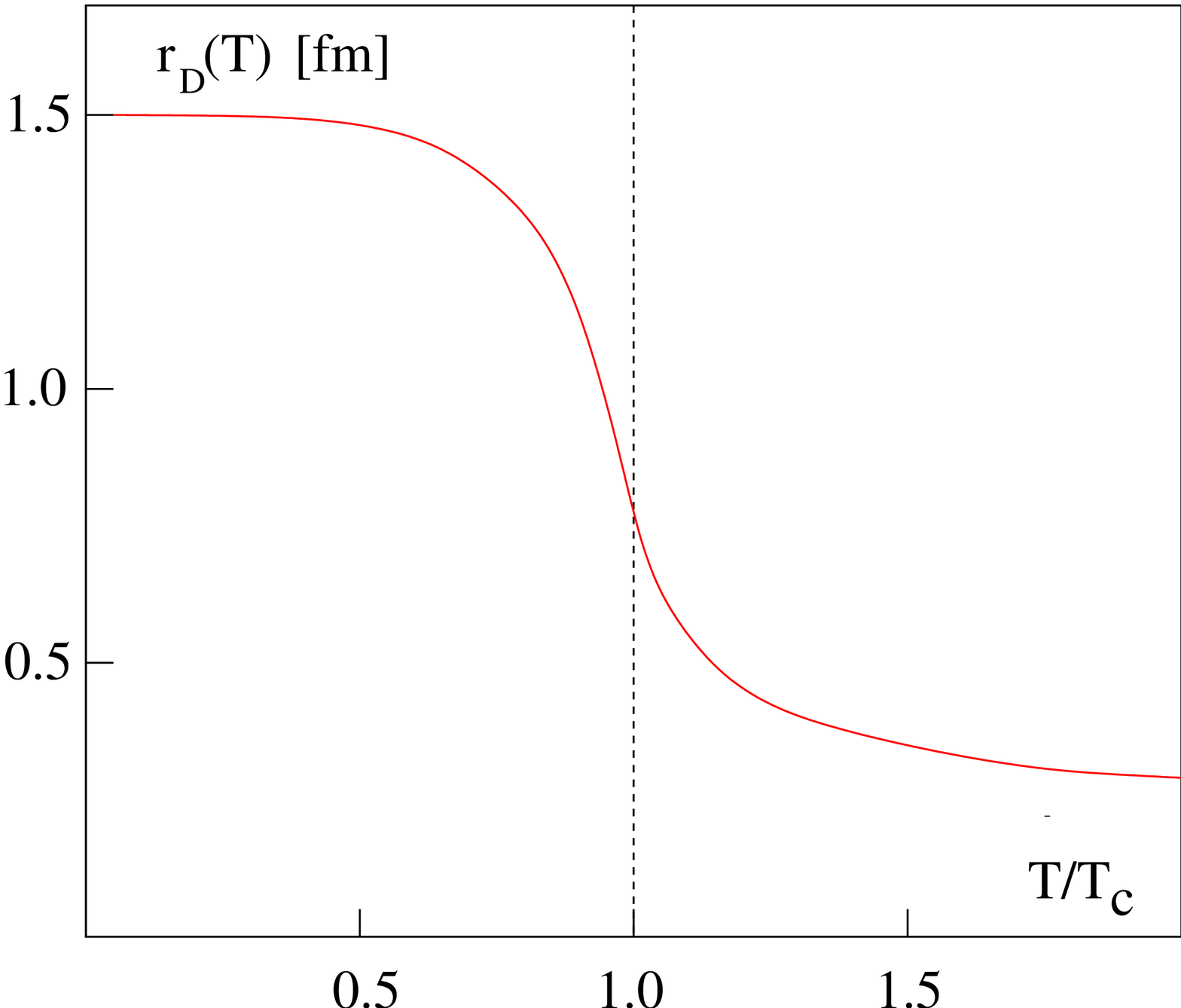,width=5cm,height=3.6cm}}
\caption{String breaking potential and interaction range at different 
temperatures}
\label{free-T}
\end{figure}

It is evident in Fig.\ \ref{OKfree} that the asymptotic value $F(\infty,T)$, 
i.e., the energy needed to separate the $\Q$ pair, decreases with increasing 
temperature, as does the separation distance at which ``the string breaks''.
For the moment we consider the latter to be defined by the point beyond which 
the free energy remains constant within errors, returning in section 4.3 to a 
more precise definition. The behaviour of both quantities is shown in Fig.\ 
\ref{free-T}. Deconfinement is thus reflected very clearly in the temperature
behaviour of the heavy quark potential: both the string breaking energy 
and the interaction range drop sharply around $T_c$. The latter decreases 
from hadronic size in the confinement region to much smaller values in the 
deconfined medium, where colour screening is operative. 

\medskip

\begin{figure}[htb]
\vspace*{-6.4cm}
\hspace*{8.5cm}
{\epsfig{file=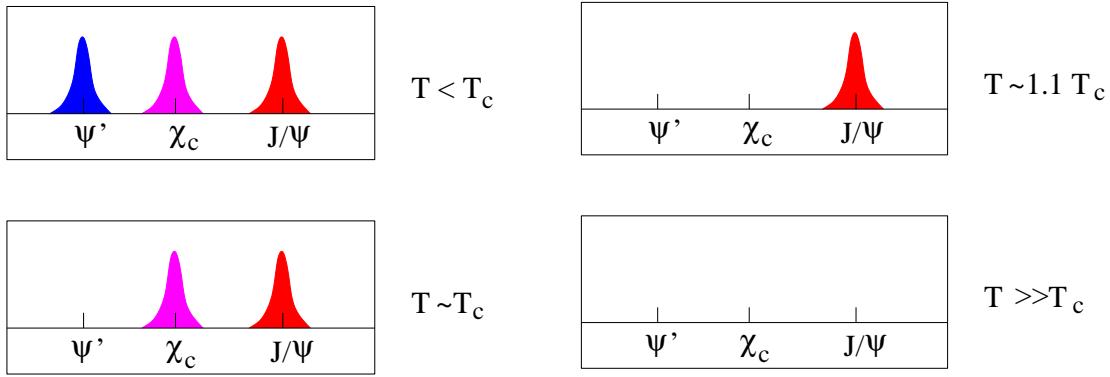,width=7cm}}
\vspace*{0.5cm}
\caption{Charmonium spectra at different temperatures}
\label{melt-t}
\end{figure}

The in-medium behaviour of heavy quark bound states thus does serve 
quite well as probe of the state of matter in QCD thermodynamics. We
had so far just considered $\Q$ bound states in general. Let us now
turn to a specific state such as the \J. What happens when the range 
of the binding force becomes smaller than the radius of the state?
Since the $c$ and the $\bar c$ can now no longer see each other, the
\J~ must dissociate for temperatures above this point. Hence the
dissociation points of the different quarkonium states provide a 
way to measure the temperature of the medium. The effect is illustrated
schematically in Fig. \ref{melt-t}, showing how with increasing 
temperature the different charmonium states ``melt'' sequentially
as function of their binding strength; the most loosely bound state 
disappears first, the ground state last.

\medskip

\begin{figure}[htb]
\hspace*{-0.3cm}
\centerline{\epsfig{file=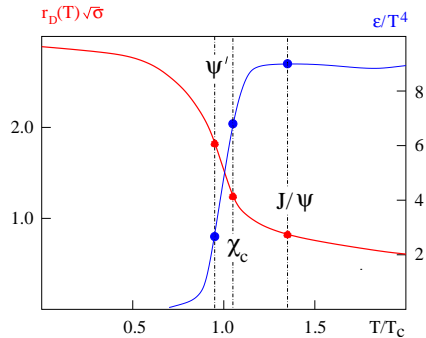,width=5.5cm}}
\caption{Charmonium dissociation vs.\ temperature and energy density}
\label{melt-e}
\end{figure}

Moreover, since finite temperature lattice QCD also provides the
temperature dependence of the energy density \cite{schlad}, the 
melting of the different charmonia or bottomonia can be specified 
as well in terms of $\e$. In Fig\ \ref{melt-e}, we illustrate this, 
combining $\e(T)$ with the force radii shown in Fig.\ \ref{free-T}. 
It is evident that although \P~and \X~are both expected to melt 
around $T_c$, the corresponding dissociation energy densities could 
still differ.

\begin{figure}[htb]
\hskip1cm
{\epsfig{file=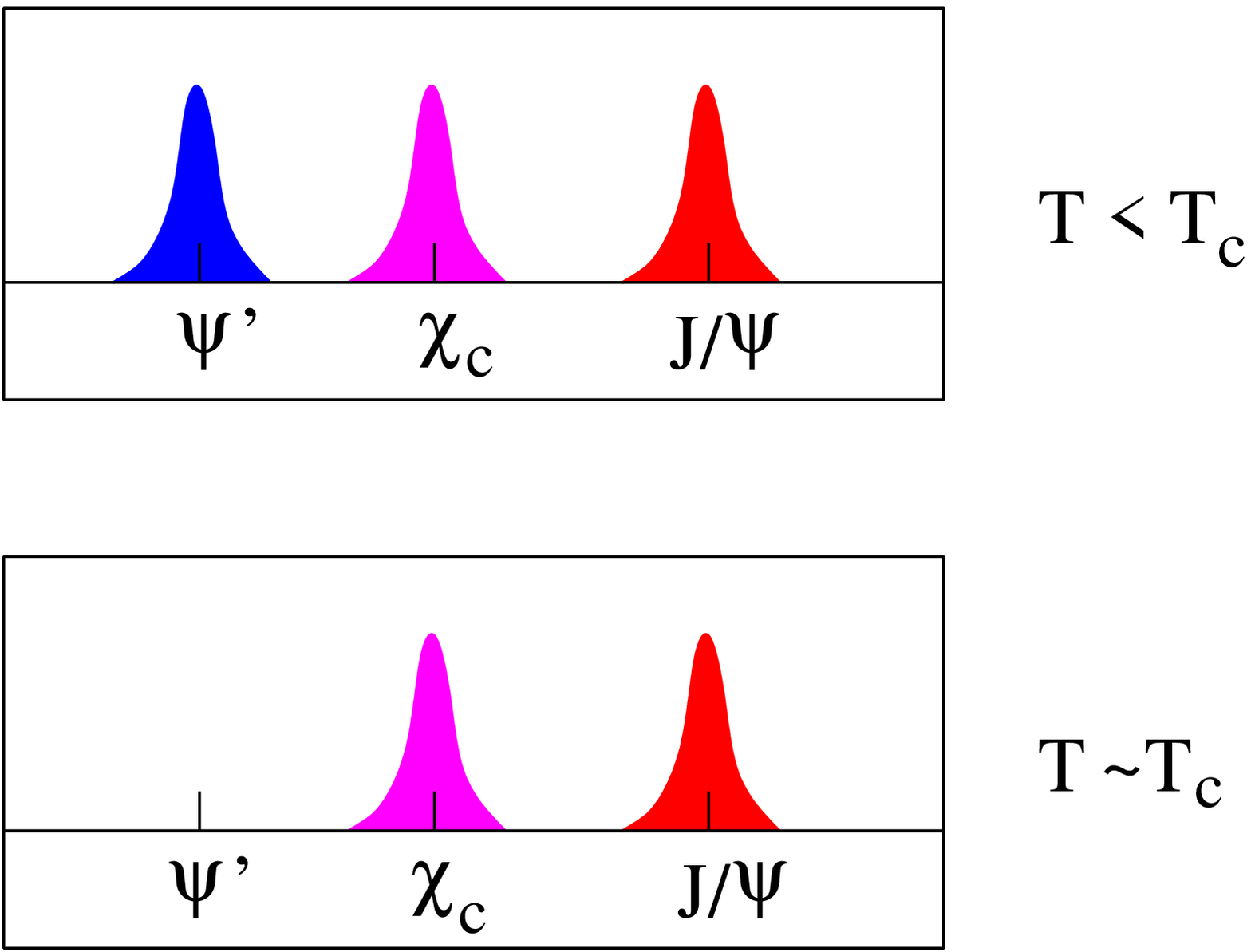,width=6.5cm}}
\end{figure}
\hskip4cm

\medskip

\vspace*{-0.3cm}

\medskip

To make these considerations quantitative, we thus have to find a way
to determine the in-medium melting points of the different quarkonium
states. This problem has been addressed in three different approaches:
\vskip-0.2cm
\begin{itemize}
\vskip-0.2cm
\item{Model the heavy quark potential as function of the temperature,
$V(r,T)$, and solve the resulting Schr\"odinger equation
(\ref{schroedinger}).}
\vspace*{-0.2cm}
\item{Determine the internal energy $U(r,T)$ of a $\Q$ pair at separation 
distance $r$ from lattice results for the corresponding free energy $F(r,T)$, 
using the thermodynamic relation 
\vskip-0.5cm
\be
U(r,T) = -T^2 \left( {\partial [F(r,T)/T] \over \partial T}\right)
= F(r,T) - T \left({\partial F(r,T) \over \partial T}\right),
\label{intern}
\ee
and solve the Schr\"odinger equation with $V(r,T)=U(r,T)$ as the binding 
potential.}
\vskip-0.5cm
\item{Calculate the quarkonium spectrum directly in finite temperature
lattice QCD.}
\end{itemize}
Clearly the last is the only model-independent way, and it will
in the long run provide the decisive determination. However, the direct
lattice study of charmonium spectra has become possible only quite
recently, and so far most results are obtained in quenched QCD 
(no dynamical light quarks);
corresponding studies for bottomonia are still more difficult.
Hence much of what is known so far is based on Schr\"odinger equation
studies with different model inputs. To illustrate the model-dependence
of the dissociation parameters, we first cite some early work using different
models for the temperature dependence of $V(r,T)$, then some recent 
studies based on lattice results for $F(r,T)$, and finally summarize
the present state of direct lattice calculations of charmonia in finite
temperature media.

\vskip0.5cm

\noindent{\large \bf 3.\ Potential Model Studies}

\bigskip

The first quantitative studies of finite temperature charmonium dissociation
\cite{K-M-S} were based on screening in the form obtained in one-dimensional 
QED, the so-called Schwinger model. The confining part of the Cornell
potential (\ref{cornell}), $V(r) \sim \sigma r$, is the solution of the 
Laplace equation in one space dimension. In this case, Debye-screening leads 
to \cite{Dixit}
\be
V(r,T) \sim \sigma r \left\{ {1-e^{-\mu r} \over \mu r} \right\}, 
\label{schwinger-e}
\ee
where $\mu(T)$ denotes the screening mass (inverse Debye radius) for the
medium at temperature $T$. This form reproduces at least qualitatively 
the convergence to a finite large distance value $V(\infty,T) = 
\sigma/\mu(T)$, and since $\mu(T)$ increases with $T$, it also gives
the expected decrease of the potential with increasing temperature.
Combining this with the usual Debye screening for the $1/r$ part of
eq.\ (\ref{cornell}) then leads to  
 \be
V(r,T) \sim \sigma r \left\{ {1-e^{-\mu r} \over \mu r} \right\} 
- {\alpha \over r} e^{-\mu r} = {\sigma \over \mu} 
\left\{ 1-e^{-\mu r} \right\} - {\alpha \over r} e^{-\mu r}
\label{schwinger}
\ee
for the screened Cornell potential. In \cite{K-M-S}, the screening mass 
was assumed to have the form $\mu(T) \simeq 4~T$, as obtained in first
lattice estimates of screening in high temperature $SU(N)$ gauge theory.  
Solving the Schr\"odinger equation with these inputs, one found that
both the \P~and the \X~become dissociated essentially at $T \simeq T_c$,
while the \J~persisted up to about $1.2~T_c$. Note that as function of
the energy density $\e \sim T^4$, this meant that the \J~really survives
up to much higher $\e$.

\medskip

This approach has two basic shortcomings:
\vspace*{-0.2cm}
\begin{itemize}
\item{The Schwinger form (\ref{schwinger-e}) corresponds to the screening
of $\sigma r$ in one space dimension; the correct result in three space
dimensions is different \cite{Dixit}.}
\vspace*{-0.2cm}
\item{The screening mass $\mu(T)$ is assumed in its high energy form;
lattice studies show today that its behaviour near $T_c$ is quite 
different \cite{D-K-K-S}.}
\end{itemize}

While the overall behaviour of this approach provides some first insight
into the problem, quantitative aspects require a more careful treatment.

\medskip

\begin{figure}[htb]
\centerline{\epsfig{file=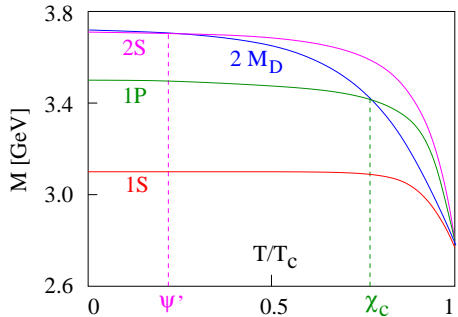,width=6cm}}
\caption{Temperature dependence of open and hidden charm masses \cite{D-P-S1}}
\label{masses}
\end{figure}

When lattice results for the heavy quark free energy as function of the 
temperature first became available, an alternative description appeared
\cite{D-P-S1}. It assumed that in the thermodynamic relation (\ref{intern})
the entropy term $-T(\partial F /\partial T)$ could be neglected, thus
equating binding potential and free energy,
\be
V(r,T) = F(r,T) -T(\partial F /\partial T) \simeq F(r,T).
\label{DPS}
\ee
Using this potential in the Schr\"odinger equation (\ref{schroedinger})
specifies the temperature dependence of the different charmonium 
masses. On the other hand, the large distance limit of $V(r,T)$ 
determines the temperature variation of the open charm meson $D$,
\be
2 M_D(T) \simeq 2 m_c + V(\infty,T)
\label{D-mass}
\ee
In fig.\ \ref{masses}, we compare the resulting open and hidden charm 
masses as function of temperature. It is seen that the \P~mass falls
below $2~M_D$ around 0.2 $T_c$, that of the \X~at about 0.8 $T_c$;
hence these states disappear by strong decay at the quoted temperatures.
Only the ground state \J~survives up to $T_c$ and perhaps
slightly above; the lattice data available at the time did not extend 
above $T_c$, so further predictions were not possible. 

The main shortcoming of this approach is quite evident. 
The neglect of the entropy term in the potential reduces $V(r,T)$ 
and hence the binding. As a result, the $D$ mass drops faster with
temperature than that of the charmonium states, and it is this effect
which leads to the early charmonium dissociation. Moreover, 
in the lattice studies used here, only the colour averaged free energy
was calculated, which leads to a further reduction of the binding
force.

\medskip

We conclude from these attempts that for a quantitative potential theory 
study, the free energy has to be formulated in the correct
three-dimensional screened Cornell form, and it then has to be
checked against the space- and temperature-dependence of the corresponding
colour singlet quantity obtained in lattice QCD. 

\vskip0.5cm

\noindent{\large \bf 4.\ Screening Theory}

\bigskip

The modification of the interaction between two charges immersed in a 
dilute medium of charged constituents is provided by Debye-H\"uckel theory, 
which for the Coulomb potential in three space dimensions leads to the 
well-known Debye screening, 
\be
{1\over r} ~\to~{1\over r} \e^{-\mu r},
\label{debye}
\ee
where $r_D=1/\mu$ defines the screening radius \cite{Landau}. 
Screening can be evaluated more generally \cite{Dixit}
for a given free energy $F(r) \sim r^q$ in $d$ space dimensions, with 
an arbitrary number $q$. We shall here apply this to the two terms
of the Cornell form, with $q=1$ for the string term, $q=-1$ for the gauge 
term, in $d=3$ space dimensions \cite{D-K-K-S}.

\medskip

We thus assume that the screening effect can be calculated separately for 
each term, so that the screened free energy becomes
\be
F(r,T) =  F_s(r,T) + F_c(r,T) =
\sigma r f_s(r,T) - {\alpha \over r} f_c(r,T).
\label{screen-cornell}
\ee
The screening functions $f_s(r,T)$ and $f_c(r,T)$ must satisfy
\begin{eqnarray}
f_s(r,T)&=&f_c(r,T)= 1 ~~{\rm for}~ T \rightarrow 0,\nonumber\\
f_s(r,T)&=&f_c(r,T)= 1 ~~{\rm for}~ r \!\rightarrow 0,
\label{screen-f}
\end{eqnarray}

since at $T=0$ there is no medium, while in the short-distance limit
$T^{-1} \gg r \to 0$, the medium has no effect. The resulting forms are
\cite{Dixit}
\be
F_c(r,T) = -{\alpha \over r}\left[e^{-\mu r} + \mu r\right]
\label{screen-gauge}
\ee
for the gauge term, and
\begin{eqnarray}
\hspace{-0.4cm}F_s(r,T) = {\sigma \over \mu}
\left[ {\Gamma\left(1/4\right) \over 2^{3/2}
\Gamma\left(3/4\right)}-{\sqrt{\mu r} \over 2^{3/4}\Gamma\left(3/4\right)}
K_{1/4}[{(\mu r)}^2]\right]
\label{screen-string}
\end{eqnarray}

for the string term. The first term in eq.\ (\ref{screen-string}) 
gives the constant large distance limit due to colour screening; the 
second provides a Gaussian cut-off in $x=\mu r$, since 
$K_{1/4}(x^2) \sim \exp\{-x^2))$; this is in contrast to the exponential 
cut-off given by the Schwinger form (\ref{schwinger}), but in accord
with string breaking arguments of Di Giacomo et al.\ \cite{A-DG}. 

\begin{figure}[htb]
\hspace*{-0.3cm}
{\epsfig{file=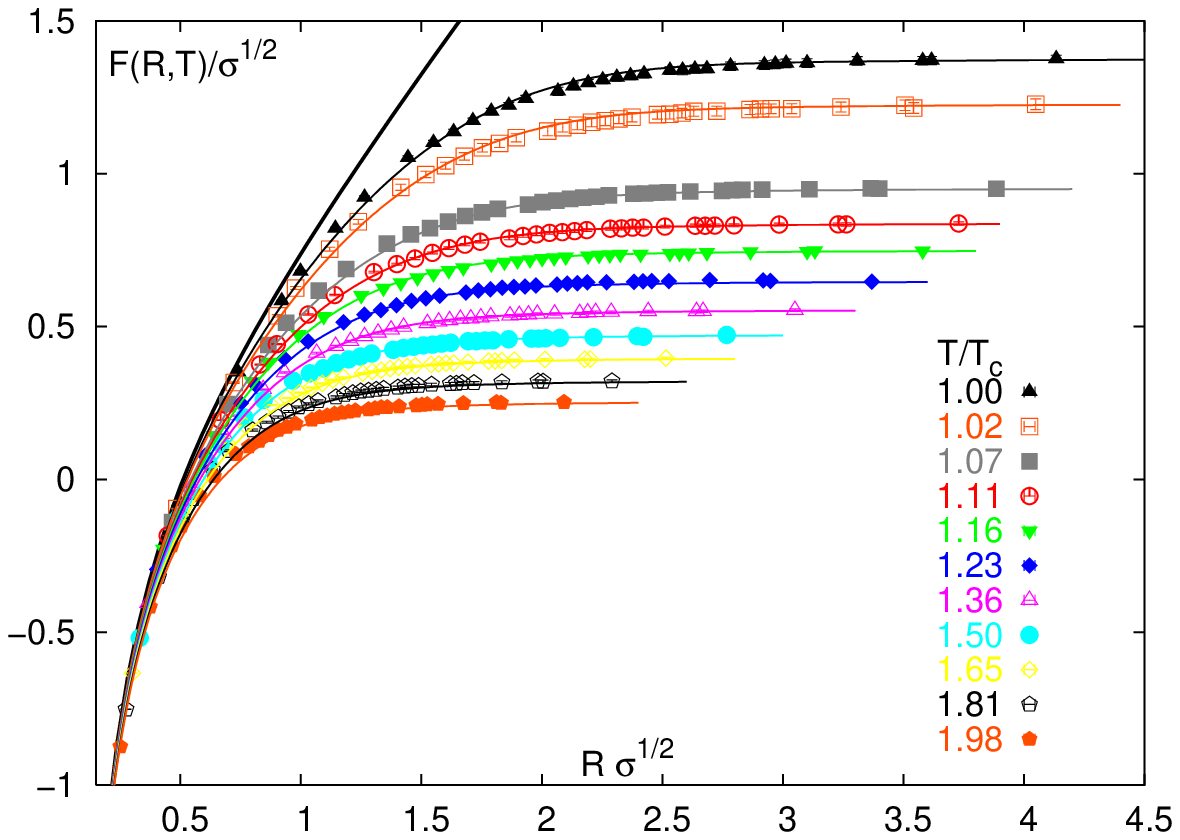,width=8cm}}
{\epsfig{file=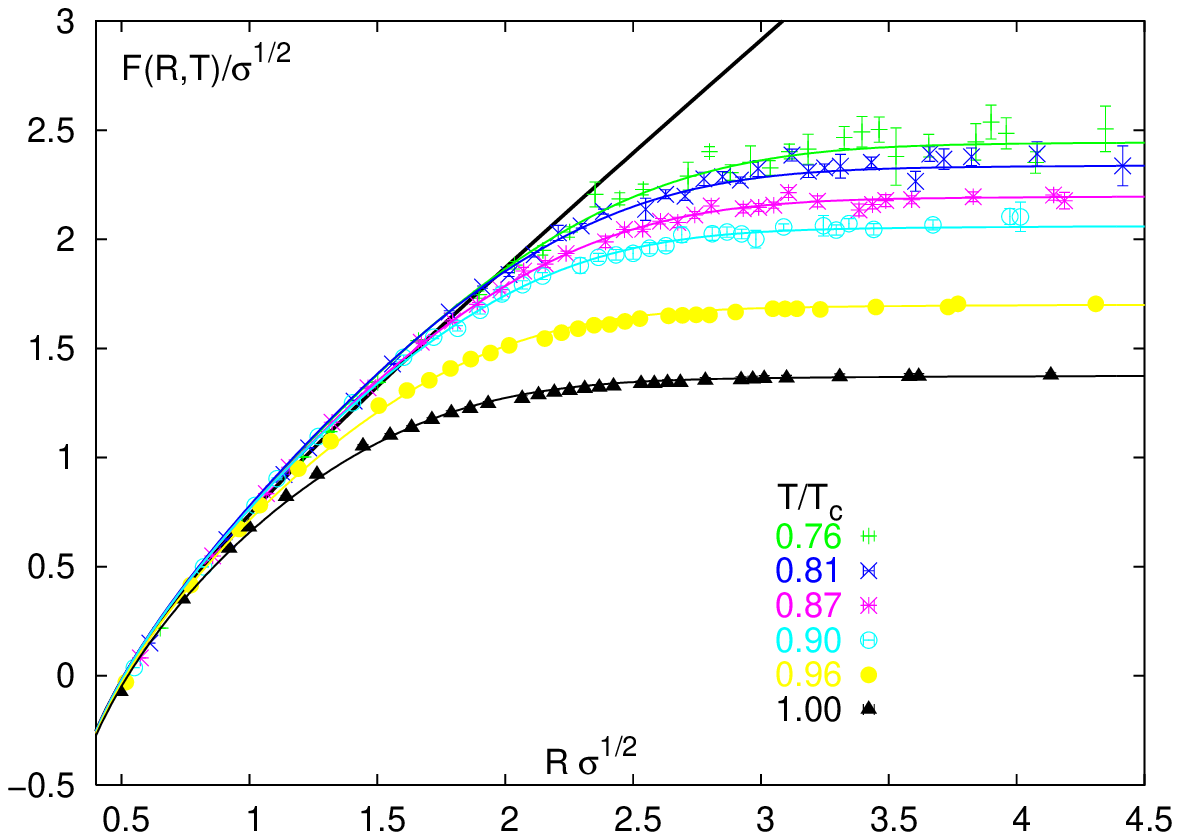,width=8cm}}
\hspace*{4cm}(a)\hspace*{7.5cm}(b)
\caption{Screening fits to the $\Q$ free energy $F(r,T)$ for $T\geq T_c$
(left) and $T\leq T_c$ (right) \cite{D-K-K-S}}
\label{free}
\end{figure}

\begin{figure}[htb]
\hspace*{-0.3cm}
\centerline{\epsfig{file=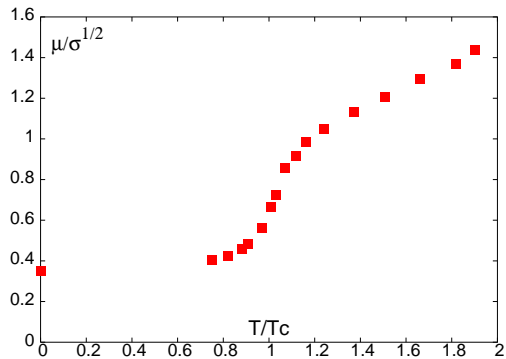,width=7cm}}
\caption{The screening mass $\mu(T)$ vs.\ $T$ \protect\cite{D-K-K-S,D-K-S}}
\label{sc-mass}
\end{figure}

At temperatures $T > T_c$, when the medium really consists of unbound
colour charges, we thus expect the free energy $F(r,T)$ to have the form
(\ref{screen-cornell}), with the two screened terms given by eqs.\ 
(\ref{screen-gauge}) and (\ref{screen-string}). In Fig.\ \ref{free}a,
it is seen that the results for the colour singlet free energy calculated
in two-flavour QCD \cite{OK} are indeed described very well by this form, 
with $m_c=1.25$ GeV and $\sqrt \sigma = 0.445$ GeV, as above.
The only parameter to be determined, the screening mass
$\mu$, is shown in  Fig.\ \ref{sc-mass}, and as expected, it first increases
rapidly in the transition region and then turns into the perturbative
form $\mu \sim T$.

\medskip

The behaviour of $\Q$ binding in a plasma of unbound quarks and gluons is
thus well described by colour screening. Such a description is in fact 
found to work well also for $T<T_c$, when quark recombination leads to an
effective screening-like reduction of the interaction range, provided one
allows higher order contributions in $x=\mu r$ in the Bessel function
$K_{1/4}(x^2)$ governing string screening \cite{D-K-K-S}. The resulting fit 
to the two-flavour colour singlet free energy below $T_c$ is shown in 
Fig.\ \ref{free}b, using
$K_{1/4}(x^2+ (1/2) x^4)$ in eq.\ (\ref{screen-string}); the corresponding 
values for the screening mass are included in Fig.\ \ref{sc-mass}.   

\medskip

With the free energy $F(r,T)$ of a heavy quark-antiquark pair given in 
terms of the screening form obtained from Debye-H\"uckel theory, the 
internal energy $U(r,T)$ can now be obtained through the thermodynamic 
relation (\ref{intern}), and this then provides the binding potential
$V(r,T)$ for the temperature dependent version of the Schr\"odinger 
equation (\ref{schroedinger}). The resulting solution then specifies
the temperature-dependence of charmonium binding as based on the 
correct heavy quark potential \cite{D-K-S}.
Let us briefly comment on the thermodynamic
basis of this approach. The pressure $P$ of a thermodynamic system is
given by the free energy, $P=-F=-U+TS$; it is determined by the kinetic
energy $TS$ at temperature $T$ and entropy $S(T)$, reduced by the potential
energy $U(T)$ between the constituents. In our case, all quantities give
the difference between a thermodynamic system containing a $\Q$ pair 
and the corresponding system without such a pair. The potential energy
of the $\Q$ pair, due both to the attraction of $Q$ and $\bar Q$ and
to the modification which the pair causes to the internal energy of 
the other constituents of the medium, is therefore given by $U$.
To determine the dissociation points for the different quarkonium
states, we thus have to solve the Schr\"odinger equation (\ref{schroedinger})
with $V(r,T)=U(r,T)$.

\medskip

\begin{figure}[h]
\begin{minipage}[t]{7cm}
\epsfig{file=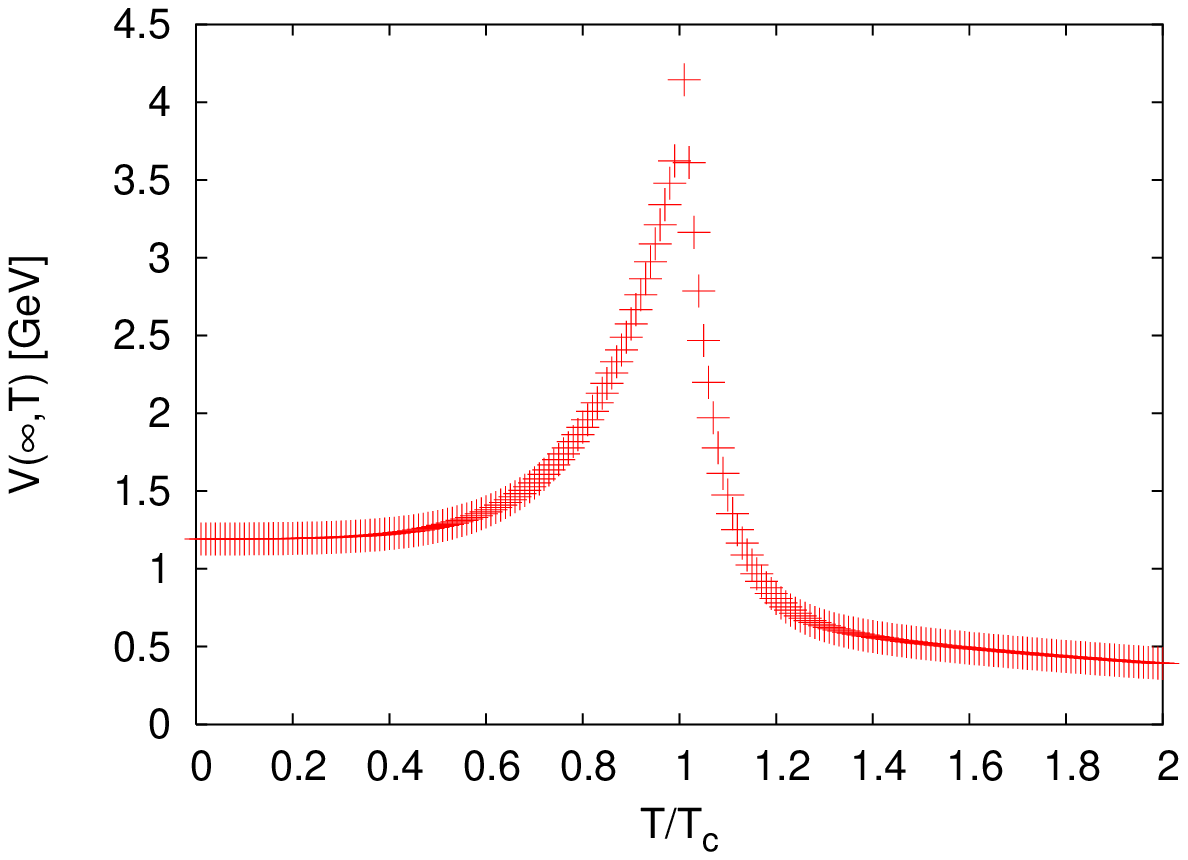,width=7.5cm}
\caption{The large distance limit of the quarkonium potential $V(r,T)$
\protect\cite{D-K-S}}
\label{vinf}
\end{minipage}
\hspace{1.3cm}
\begin{minipage}[t]{7cm}
\vspace*{-5.3cm}
\hskip-0.3cm
\epsfig{file=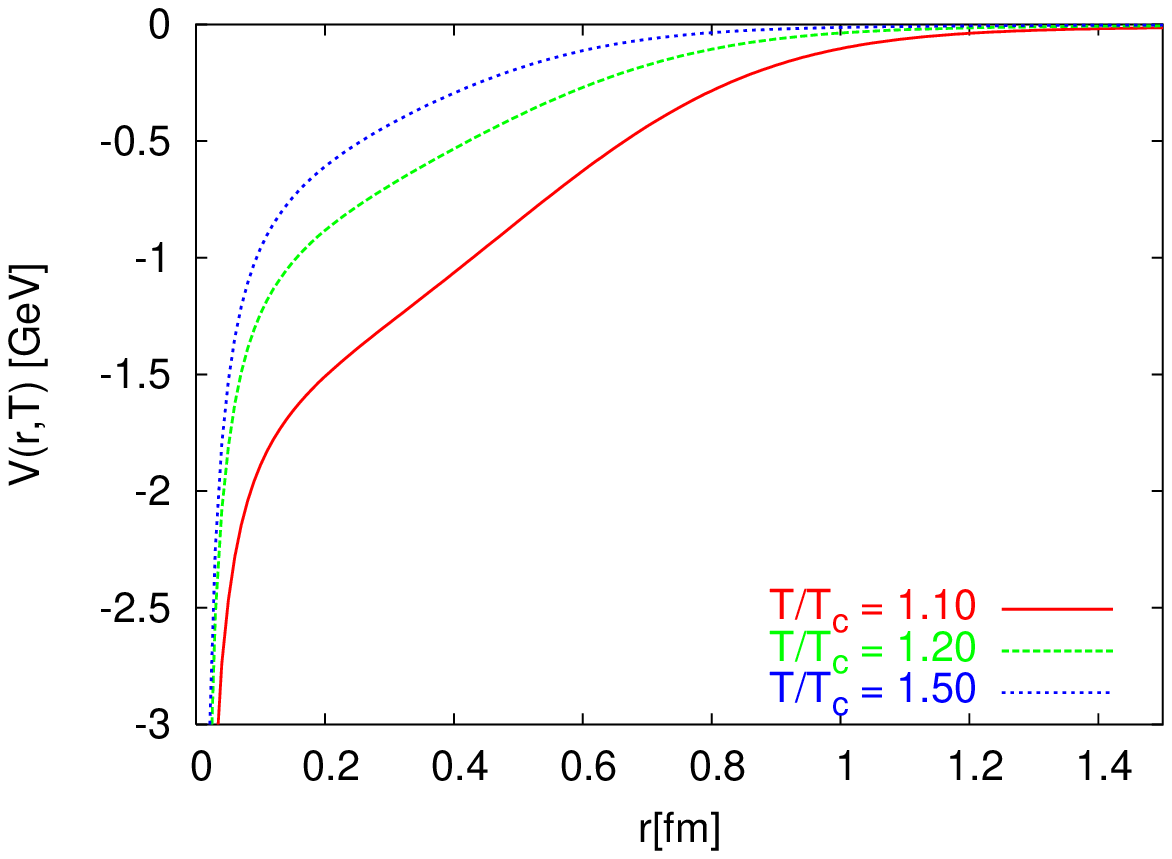,width=7.5cm}
\caption{Variation of $\tilde V(r,T)$ with $r$ for different $T$
\protect\cite{D-K-S}}
\label{vr}
\end{minipage}
\end{figure}

From eqs.\ (\ref{screen-gauge}) and (\ref{screen-string}) we 
get for the $\Q$ potential
\be
V(r,T)= V(\infty,T) + \tilde V(r,T),
\ee
with 
\be
V(\infty,T) = c_1{\sigma \over \mu} - \alpha \mu
+ T{d\mu \over dT}[c_1 {\sigma \over \mu^2} + \alpha],
\ee
and $c_1=\Gamma(1/4)/2^{3/2} \Gamma(3/4)$, and where $\tilde V(r,T)$ 
contains the part of the potential which vanishes for $r\to \infty$.
The behaviour of $V(\infty,T)$ as function of the temperature is
shown in Fig.\ \ref{vinf}. It measures (twice) the energy of the cloud
of light quarks and gluons around an isolated heavy quark, of an extension 
determined by the screening radius, relative to the energy contained in
such a cloud of the same size without a heavy quark. This energy difference 
arises from the interaction of the heavy quark with light quarks and gluons
of the medium, and from the modification of the interaction between the light 
constituents themselves, caused by the presence of the heavy charge. --
The behaviour of $\tilde V(r,T)$ is shown in Fig.\ \ref{vr} for three
different values of the temperature. It is seen that with increasing $T$,
screening reduces the range of the potential. 

\medskip

The relevant Schr\"odinger equation now becomes
\be
\left\{{1\over m_c}\nabla^2 - \tilde V(r,T)\right\} \Phi_i(r) = 
\Delta E_i(T) \Phi_i(r)
\label{T-schroedinger}
\ee 
where
\be
\Delta E_i(T)  = V(\infty,T) - M_i - 2m_c
\ee
is the binding energy of charmonium state $i$ at temperature $T$.
When it vanishes, the bound state $i$ no longer exists, so that 
$\Delta E_i(T)=0 $ determines the dissociation temperature $T_i$ 
for that state. The temperature enters only through the $T$-dependence
of the screening mass $\mu(T)$, as obtained from the analysis of the 
lattice results for $F(r,T)$. -- Equivalently, the divergence of the
binding radii, $r_i(T) \to \infty$, can be used to define the 
dissociation points $T_i$.

\medskip

The same formalism, with $m_b=4.65$ GeV replacing $m_c$, leads to the
bottomonium dissociation points. The combined quarkonium results are 
listed in Table 2. They agree quite well with those obtained in a
very similar study based on corresponding free energies obtained
in quenched lattice QCD \cite{Wong}, indicating that above $T_c$ gluonic 
effects dominate. Using a parametrically generalized screened Coulomb
potential obtained from lattice QCD results also leads to very compatible 
results for the $N_f=2$ and quenched cases\cite{Alberico}. We recall
here that the main underlying change, which is responsible for the much
higher dissociation temperatures for the quarkonium ground states, is the 
use of the full internal energy (\ref{intern}), including the entropy term,
as potential in the Schr\"odinger equation: this makes the binding much
stronger. 

\medskip

We should note, however, that in all such potential studies
it is not so clear what binding energies of less than a few MeV 
or bound state radii of several fermi can mean in a medium whose
temperature is above 200 MeV and which leads to screening radii 
of less than 0.5 fm. In such a situation, thermal activation \cite{K-ML-S}
can easily dissciate the bound state.

\medskip

\begin{center}
\renewcommand{\arraystretch}{2.0}
\begin{tabular}{|c||c|c|c||c|c|c|c|c|}
\hline
 state & J/$\psi(1S)$ & $\chi_c$(1P) & $\psi^\prime(2S)$&$\Upsilon(1S)$&
$\chi_b(1P)$&$\Upsilon(2S)$&$\chi_b(2P)$&$\Upsilon(3S)$\\
\hline
\hline
$T_d/T_c$ & 2.10  & 1.16 & 1.12 & $>4.0$ & 1.76 & 1.60& 1.19 & 1.17 \\
\hline
\end{tabular}\end{center}

\medskip

\centerline{Table 2: Quarkonium Dissociation Temperatures \cite{D-K-S}}

\vskip0.5cm

\noindent{\large \bf 5.\ Charmonium Correlators}

\bigskip

\begin{figure}[htb]
\centerline{\epsfig{file=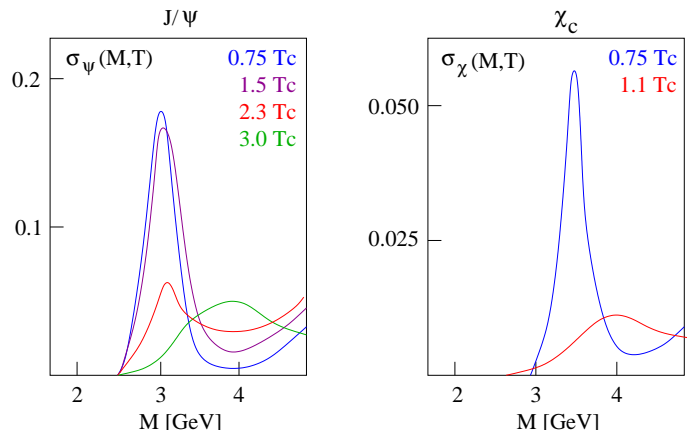,width=9cm}}
\caption{\J~and \X~spectral functions at different temperatures}
\label{spec}
\end{figure}

The direct spectral analysis of charmonia in finite temperature lattice 
has come within reach only in very recent years \cite{lattice-charm}.
It is possible now to evaluate the correlation functions $G_H(\t,T)$ for
hadronic quantum number channels $H$ in terms of the Euclidean time
$\t$ and the temperature $T$. These correlation functions are directly
related to the corresponding spectral function $\sigma_H(M,T)$,
which describe the distribution in mass $M$ at temperature $T$ for
the channel in question. In Fig.\ \ref{spec}, schematic results at different
temperatures are shown for the \J~and the \X~channels. It is seen that
the spectrum for the ground state \J~ remains essentially unchanged
even at $1.5~T_c$. At $3~T_c$, however, it has disappeared; the
remaining spectrum is that of the $\C$ continuum of \J~quantum numbers
at that temperature. In contrast, the \X~is already absent at $1.1~T_c$, 
with only the corresponding continuum present. 

\medskip

These results are clearly very promising: they show that in a foreseeable
future, the dissociation parameters of quarkonia can be determined 
{\sl ab initio} in lattice QCD. For the moment, however, they remain
indicative only, since the underlying calculations were generally
performed in quenched QCD, i.e., without dynamical quark loops. Since 
such loops are crucial in the break-up of quarkonia into light-heavy mesons, 
final results require calculations in full QCD. Some first calculations
in two-flavour QCD have just appeared \cite{skullerud} and support
the late dissociation of the \J.  The widths of the
observed spectral signals are at present determined by the precision of
the lattice calculations; to study the actual physical widths, much
higher precision is needed. Finally, one has so far only first signals
at a few selected points; a temperature scan also requires higher performance
computational facilities. Since the next generation of computers, in the
multi-Teraflops range, is presently going into operation, the next
years should bring the desired results. So far, in view of the mentioned
uncertainties in both approaches, the results from direct lattice studies
and those from the potential model calculations of the previous section
appear quite compatible. 

\vskip0.5cm

\noindent{\large \bf 6.\ Conclusions}

\bigskip

The theoretical analysis of the in-medium behaviour of quarkonia has 
greatly advanced in the past decade. Potential model studies based on
lattice results for the colour-singlet free energy appear to be converging
to results from direct lattice calculations, and within a few more
years, the dissociation temperatures for the different quarkonium
states should be known precisely. Through corresponding calculations
of the QCD equation of state, these temperatures provide the energy
density values at which the dissociation occurs. In statistical QCD,
quarkonia thus allow a spectral analysis of the quark-gluon plasma.
as many of the observed features as possible. 

\medskip

Finally, it seems worthwhile to note that experimental measurements of 
the relative dissociation points of the different quarkonium states could
be possible in the study of high energy nucleus-nucleus collisions. If
they succeed, this might in fact be a unique chance to 
compare quantitative {\sl ab initio} QCD predictions directly to data. 

\vskip1cm

\centerline{\large \bf Acknowledgements}

\bigskip

It is a pleasure to thank S.\ Digal, O.\ Kaczmarek,
F.\ Karsch, D.\ Kharzeev, P.\ Petreczky, 
M.\ Nardi and R.\ Vogt for stimulating and helpful comments.

\end{document}